\documentclass{ws-ijqi}
\usepackage{epsfig,graphicx,bm}

\usepackage{amsmath}
\usepackage{latexsym}
\usepackage{amsfonts}
\usepackage{amssymb}

\DeclareMathOperator{\Tr}{tr}
\DeclareMathOperator{\Id}{ {\bf 1}}

\newcommand{\beq}{\begin{equation}}
\newcommand{\eeq}{\end{equation}}
\newcommand{\beqa}{\begin{eqnarray}}
\newcommand{\eeqa}{\end{eqnarray}}

\begin{document} 

\markboth{Dagmar Bru{\ss} \emph{et al.}}
{Dense coding with multipartite quantum states}

\catchline{}{}{}{}{}

\title{Dense coding with multipartite quantum states}

\author{Dagmar Bru{\ss}}

\address{Institut f\"ur Theoretische Physik III, Heinrich-Heine-Universit\"at 
D\"usseldorf, D-40225 D\"usseldorf, Germany}

\author{Maciej Lewenstein\footnote{Also at Instituci\'o Catalana de Recerca i Estudis Avan{\c c}ats.}, Aditi Sen(De), Ujjwal Sen}

\address{ICFO-Institut de Ci\`encies Fot\`oniques, Jordi Girona 29, 
Edifici Nexus II, E-08034 Barcelona, Spain, and \\
Institut f\"ur Theoretische Physik, 
Universit\"at Hannover, D-30167 Hannover,
Germany}

\author{Giacomo Mauro D'Ariano and 
Chiara Macchiavello}

\address{Dipartimento di Fisica ``A. Volta" and INFM-Unit\'a di Pavia,
Via Bassi 6, I-27100 Pavia, Italy}

\maketitle

\begin{abstract}

We consider generalisations of the dense coding protocol with an 
arbitrary number of senders and 
either one or two receivers, sharing a multiparty quantum state, and using a 
noiseless channel. 
For the case of a single receiver, the capacity of such information transfer 
is found exactly. It is 
shown that  the capacity is not enhanced by allowing 
the senders to perform joint operations. 
We provide a nontrivial upper bound on the capacity in the case of two 
receivers.
We also give a classification of the set of all multiparty states in terms of 
their usefulness for dense coding. 
We provide examples for each of these classes, and discuss some of their 
properties.

\end{abstract}



\section{Introduction}

Entanglement among quantum systems can be used to perform tasks that are not 
possible with classical states. 
Phenomena where entanglement plays a crucial role include e.g. teleportation 
\cite{BBCJPW} and  dense coding \cite{nischintapur}.
In the dense coding protocol, entangled quantum states are used to send 
classical information from a sender (say, Alice) to a receiver (say, Bob).
Suppose that Alice wants to send two bits of classical information to Bob.  
Then the Holevo bound, to be discussed later, shows that 
Alice must send two qubits (two-dimensional quantum states) to Bob, if only a 
noiseless quantum channel 
is available. However, if Alice and Bob have previously shared entanglement, 
then Alice may have to send less than two  qubits to Bob. It was shown by Bennett and Wiesner \cite{nischintapur}, that 
by using a previously shared singlet, Alice will be able to send two bits to Bob, by transmitting just a single qubit. 

To consider a realistic scenario, two avenues are usually taken. 
One approach is to consider a \emph{noisy} quantum channel, where the 
additional resource is an arbitrary amount of shared bipartite \emph{pure}
state entanglement (see e.g. \cite{Bennett-ek,Bennett-dui,MarieCurie,Debu}).
This is the scenario  of
 the so-called entanglement assisted capacity, which  
refers to a property of the channel. 
The other approach is to consider a \emph{noiseless} quantum channel, while 
the assistance is by a given bipartite \emph{mixed} entangled state 
(see e.g. \cite{MarieCurie,Debu,ek,dui,tin,char}). In this second case
the capacity refers to a feature of the state. 
In this paper, we consider the second approach, 
in the general situation of several senders and one or two receivers. 
Therefore the  senders and the receiver(s) share a given multiparty 
state. 
The senders (called Alices, and named as \(A_1,A_2, \cdots, A_N\)) want to send classical information to the receivers (Bobs, \(B_1\) and \(B_2\)), where
the information of one Alice can be different from that of another. All the parties that take part in the protocol are 
at distant locations. Consequently, both the encoding of the information by the Alices, and the   decoding of  
it by the Bobs, must be by local operations. Additionally, the Alices can communicate  between themselves over a classical 
channel, and likewise the Bobs can do so between themselves. Classical communication is of course not allowed between the senders and the 
receivers.

We considered this scenario in Ref. \cite{TajMahal}, and named it  
``distributed quantum dense coding''. In this paper, we further discuss 
the bounds on the capacity of dense coding 
in this scenario, for a given state, where the 
capacity is defined as the number of classical bits that can be accessed 
by the receivers, per use of the noiseless channel.
Also, we give a classification of multipartite states according to their degree of ability to 
assist in distributed dense coding.

The paper is organized as follows.
In Section \ref{sec-halum} we discuss the Holevo bound, which is
a crucial element in finding the capacity of dense coding for the case of a single receiver.
In Section \ref{koto-capa-re?}, we consider the case of dense coding with a single sender and a single receiver. 
In Section \ref{Ujjain}, we take up the case of many senders but a single receiver, and find the capacity in this scenario. 
We show that the capacity is not enhanced by allowing the 
senders to perform joint operations. 
To consider the case of many receivers, we must obtain a Holevo-like upper bound on classical information that can be decoded from 
multiparty quantum ensembles. Such a bound, derived  in Ref. \cite{Khajuraho} 
for bipartite ensembles, is discussed in Section \ref{sec-RoyalBengal}.
In Sec. \ref{sec-dui-receivers} we obtain an upper  
bound of dense coding schemes 
for an arbitrary number of senders and two receivers (a bound for  
multiparty ensembles is currently absent \cite{Sundarban}).  
In Sec. \ref{sec-nyapla}, we will discuss a classification of multiparty states according to their 
degree of usefulness in dense coding protocols and give some examples.
In Sec. \ref{disco} we will summarize our results and discuss some 
related open 
problems.

\section{The Holevo bound}
\label{sec-halum}

The Holevo bound is an upper bound on the amount of classical information that 
can be accessed from a quantum ensemble in which the information is encoded.
Suppose that Alice (\(A\)) has the classical message \(i\) that 
occurs with probability \(p_i\). Alice encodes this information \(i\) in a quantum state \(\rho_i\), and sends it to Bob. 
Bob receives the ensemble \(\{p_i, \rho_i\}\), and wants to obtain as much information as possible about \(i\). To do so, 
he performs a measurement, that gives the result \(m\) with probability \(q_m\). Let the corresponding post-measurement ensemble be 
\(\{p_{i|m}, \rho_{i|m}\}\). The information gathered can be quantified by the mutual information between the 
message index \(i\) and the measurement outcome \cite{Chennai}:
\beq
I(i:m)= H(\{p_i\}) - \sum_m q_m H(\{p_{i|m}\}).
\eeq

Here \(H(\{r_x\}) = -\sum_xr_x\log_2r_x\) is the Shannon entropy of the probability distribution \(\{r_x\}\).
Bob will be interested to obtain the maximal information, which is maximum of \(I(i:m)\) for all measurement strategies. This 
quantity is called the accessible information:
\beq
I_{acc} = \max I(i:m),
\eeq 
where the maximization is performed over all measurement strategies. 

The maximization involved in the definition of accessible information is usually hard to compute, and hence the importance of bounds \cite{ref-halum,Utpakhi}. 
In particular, in Ref. \cite{ref-halum}, a universal upper bound on 
\(I_{acc}\), the Holevo bound,  is given (see also 
\cite{Rajabazar1,Khajuraho,Rajabazar2})
\beq
I_{acc}(\{p_i, \rho_i\}) \leq \chi(\{p_i, \rho_i\}) \equiv S(\overline{\rho}) - \sum_i p_i S(\rho_i).
\eeq
Here \(\overline{\rho} = \sum_ip_i\rho_i\) is the average ensemble state, and 
\(S(\varsigma)= - \mbox{tr}(\varsigma \log_2 \varsigma)\) 
is the von Neumann entropy of \(\varsigma\). 
The Holevo bound is asymptotically  achievable in the sense that if the sender 
is able to send long strings of the input quantum 
states \(\rho_i\), then there exists a particular encoding and a decoding scheme that asymptotically attains the bound \cite{babarey}.

\section{Capacity of dense coding with one sender and one receiver}
\label{koto-capa-re?}

Suppose that Alice and Bob share a quantum state \(\rho^{AB}\). Alice performs the unitary operation \(U_i\) with probability \(p_i\), on her 
part of the state \(\rho^{AB}\) to encode the classical information  \(i\). 
Subsequent to her unitary rotation, she sends her part of the state 
\(\rho^{AB}\) to Bob. Bob then has the ensemble \(\{p_i, \rho_i\}\), 
where 
\begin{equation}
\rho_i = U_i \otimes \Id \rho^{AB} U_i^{\dagger} \otimes \Id.
\end{equation}

The information that Bob is able to gather is \(I_{acc}(\{p_i, \rho_i\})\). This quantity is bounded from above by 
\(\chi(\{p_i, \rho_i\})\). 
The ``one-capacity'' \(C^{(1)}\) of dense coding for the state \(\rho^{AB}\) is the Holevo bound for the best encoding by Alice:
\begin{equation}
\label{Kohinoor}
C^{(1)}(\rho) = \max_{p_i,U_i} \chi(\{p_i, \rho_i\}) \equiv \max_{p_i,U_i} \left(S(\overline{\rho}) - \sum_i p_i S(\rho_i)\right).
\end{equation}
The superscript \((1)\) reflects the fact that Alice is using the shared 
state once at a time, during the asymptotic process. 
She is not using entangled unitaries on more than one copy of her parts of the 
shared states \(\rho^{AB}\). As we will see below, encoding with entangled 
unitaries does not help her to send more information to Bob. 

In performing the maximization in Eq. (\ref{Kohinoor}), first note that the second term in the right hand side (rhs) is equal to 
\(-S(\rho)\), for all choices of the unitaries and probabilities, as
unitary operations do not change the spectrum, and hence 
the entropy, of a state. Secondly, we have 
\beq
S(\overline{\rho}) \leq S(\overline{\rho}^A) 
+ S(\overline{\rho}^B) \leq \log_2 d_A + S(\overline{\rho}^B),
\eeq 
where \(d_A\) is the dimension of Alice's part of the Hilbert space of 
\(\rho^{AB}\), and 
 \( \overline{\rho}^A = \mbox{tr}_B \overline{\rho}\), \( \overline{\rho}^B = \mbox{tr}_A \overline{\rho}\). 
Moreover, \(S(\overline{\rho}^B) = S(\rho^B)\),
as nothing was done at Bob's end during the encoding procedure. Therefore, 
we have 
\beq
\max_{p_i,U_i} S(\overline{\rho}) \leq \log_2 d_A + S(\rho^B).
\eeq

This bound is reached by any 
complete set of orthogonal unitary operators
$\{W_j\}$, to be  chosen with equal probabilities, 
which satisfy the {\em trace rule} 
$\frac{1}{d_A}\sum_{j=1}^{d_A} W_j^\dag \Xi W_j=\Tr[\Xi]I$, 
for any operator $\Xi$. Therefore, we have 
\beq
C^{(1)}(\rho) = \log_2 d_A + S( \rho^B ) - S(\rho). 
\eeq

The optimization procedure above essentially follows that in Ref. \cite{tin}.
Several other lines of argument are possible for the maximization. 
One approach is given in Ref. \cite{dui} (see also \cite{TajMahal}). 
Another way to proceed is 
to guess where the maximum is reached, 
and then perturb the guessed result. If the 
first order perturbations vanish, the guessed result is correct, 
as the von Neumann entropy is a concave function and 
the maximization is carried out over the continuous set of all 
\(\{p_i, U_i\}\) \cite{TajMahal}. 
Note here that without using the additional resource of entangled states, 
Alice will be able to reach a capacity of just \(\log_2d_A\) bits. 
Therefore, entanglement in a state \(\rho^{AB}\) is useful for dense 
coding if \(S( \rho^B ) - S(\rho)>0\). 
Such states exist, an example being the singlet state.

\subsection{Entangled encoding and the asymptotic capacity}
\label{kochi-result}

Suppose now that Alice is able to use entangled unitaries on two copies of the shared state \(\rho\). For definiteness, let us call the copies 
\(\rho^{a_1b_1}\) and \(\rho^{a_2b_2}\) (\(a_1\) and \(a_2\) refer to Alice's states, \(b_1\) and \(b_2\) to Bob's). 
Alice may possibly apply unitaries \(U_i\)  that cannot 
be written as \(U_i = U_i^{a_1} \otimes U_i^{a_2}\). 
Applying such a general set of unitaries \(U_i\) with 
probabilities \(p_i\), the output ensemble is \(\{p_i,\rho_i^{(2)}\}\), where  
\(\rho_i^{(2)} = U_i^{a_1a_2} \otimes \Id \otimes \Id \left(\rho^{a_1b_1} \otimes \rho^{a_2b_2}\right) U_i^{a_1a_2 \dagger} \otimes \Id \otimes \Id\).
It is natural to define the ``two-capacity'' of dense coding for the state \(\rho\) as
\begin{eqnarray}
\label{Eli-aur-kaa!}
C^{(2)}(\rho) = \frac{1}{2} \max_{p_i,U_i} \chi(\{p_i, \rho_i^{(2)}\}) 
\equiv\frac{1}{2} \max_{p_i,U_i} \left(S(\overline{\rho}^{(2)}) 
- \sum_i p_i S(\rho_i^{(2)})\right),
\end{eqnarray}
where 
\(\overline{\rho}^{(2)} = \sum_i p_i \rho_i^{(2)}\). 
Again the second term within the maximization of Eq. (\ref{Eli-aur-kaa!}) 
is just \(-S(\rho \otimes \rho) = -2S(\rho)\). 
The first term is bounded from above by 
\(\log_2 (d_{a_1} d_{a_2}) + S(\rho^{b_1} \otimes \rho^{b_2}) = 2\log_2 d_A + 2 S(\rho^B)\), which can be reached by any complete set of orthogonal unitaries 
on \(A_1A_2\) that satisfies the trace rule. (Here \(d_{a_j}\) 
is the dimension of the particle 
\(a_j\), and \(\rho^{b_j} = \mbox{tr}_{a_j} \rho^{a_j b_j}\), where \(j=1,2\).) However, one such
set of unitaries is formed by tensor products of two complete sets of orthogonal unitaries on \(A_1\) and \(A_2\). Therefore, product
unitaries are enough to  attain \(C^{(2)}\), and its value is equal to that of \(C^{(1)}\). Similar arguments hold for 
\(C^{(L)} (\rho) = \frac{1}{L} \max_{p_i,U_i} \chi(\{p_i, \rho_i^{(L)}\})\) 
for any \(L\), where the \(U_i\)'s are now possibly entangled unitaries over the \(L\)-fold tensor product 
of the Hilbert space on Alice's side. Consequently, the asymptotic capacity 
(henceforth called capacity) 
of dense coding of a bipartite state \(\rho^{AB}\) is given by
\begin{equation}
\label{ek-ekcapacity}
C(\rho) = \lim_{L \rightarrow \infty} C^{(L)}(\rho) = \log_2 d_A + S( \rho^B ) - S(\rho). 
\end{equation}

Note however that this additivity is shown only in the case of encoding by unitary operations. 
In this paper, both in the bipartite as well as in the multipartite scenario,
we will consider unitary encoding only.

\subsection{Bipartite bound entangled states}
\label{sec-Leibnizhaus}

A bipartite state \(\rho^{AB}\) is useful for dense coding if and only if 
\(S( \rho^B ) - S(\rho)>0\). We now show that this relation cannot hold for bipartite bound entangled states \cite{ref-Gdansk}.  
Let us first state the reduction criterion  \cite{reduction} 
for detecting distillable states: 
If a state \(\rho^{AB}\) is separable or bound entangled, then 
\(\rho^A \otimes I_{d_B} \geq \rho^{AB}\) 
and \(I_{d_A} \otimes \rho^B \geq \rho^{AB}\). There exist
 distillable states that violate this criterion. 
Any state \(\rho^{AB}\) for which 
\(S(\rho^B) - S(\rho^{AB}) >0\) violates the 
reduction criterion \cite{Wolf} (see also \cite{etao-dewa-uchit}), and is
hence distillable. Therefore, a state that is useful for dense coding is 
always distillable, i.e. free entangled.  
It has been shown that bound entangled states are not useful for sending classical information even by more general 
encoding operations \cite{MarieCurie}.

\section{Capacity of dense coding with many senders and one receiver}
\label{Ujjain}

Suppose now that there are \(N\) Alices, viz. \(A_1\), \(A_2\), \(\cdots\), \(A_N\), who want to send information to a single receiver,
Bob (\(B\)). They share the quantum state \(\rho^{A_1A_2 \cdots A_N B}\). Depending on the classical information \(i_k\) that 
\(A_k\) wants to send to Bob, she applies the unitary operation \(U_{i_k}\) with probability \(p_{i_k}\) (\(k = 1,2, \cdots, N\)). 
After applying the unitary operations, they send their parts of the quantum state to Bob, who has now the ensemble 
\(\{p_{\{i\}}, \rho_{\{i\}}\}\), where \(\{i\}\) denotes the string \(\{i_1,i_2,\cdots,i_N\}\). Moreover
\begin{eqnarray}
p_{\{i\}} = p_{i_1}p_{i_2} \cdots p_{i_N},\qquad
\rho_{\{i\}} =  U_{\{i\}} \otimes  \Id \rho^{A_1A_2 \cdots A_N B} 
U_{\{i\}}^\dagger \otimes \Id,
\end{eqnarray} 
where $U_{\{i\}} = U_{i_1} \otimes U_{i_2} \otimes \cdots \otimes U_{i_N}$.
The task of Bob is to obtain as much information  as 
possible about the message string \(\{i\}\). 
Since the Holevo bound is asymptotically attainable by product encoding (Section \ref{sec-halum}),  
the ``one-capacity'' of the state \(\rho^{A_1A_2 \cdots A_N B}\) in this case is defined as 
\beq
C^{(1)}(\rho) = \max_{p_{\{i\}}, U_{\{i\}}} \chi(\{p_{\{i\}}, \rho_{\{i\}}\}).
\eeq

To avoid multiple indices, 
we use the same notation as in the case of a single sender. As we will 
see, the capacities in the case of a single sender and multiple senders are the same (at least in the case when there is 
only a single receiver).
Analogous considerations as for the maximization of Eq. (\ref{Kohinoor}) 
lead to 
\beq
C^{(1)}(\rho) = \log_2 d_{A_1} + \log_2 d_{A_2} + \cdots + \log_2 d_{A_N}  + S( \rho^B ) - S(\rho),
\eeq
where \(d_{A_k}\) is the dimension of the Hilbert space in possession of the \(k\)th Alice \(A_k\). 
Moreover by similar arguments as in Section \ref{kochi-result}, also in this 
case, the one-capacity can be shown to be the 
asymptotic capacity, so that 
\begin{equation}
\label{kanishka}
C(\rho) = \log_2 d_{A_1} + \log_2 d_{A_2} + \cdots + \log_2 d_{A_N}  + S( \rho^B ) - S(\rho).
\end{equation}

Again, we use the same notation as in the case of a single sender.
The capacity is reached by any complete set of orthogonal unitaries that satisfies the trace rule. However
such a complete orthogonal set of unitaries of the \(A_1 A_2 \cdots A_N\) space can be formed by \emph{product} unitaries of the 
individual spaces of the \(A_k\). This leads us to the 
conclusion that even if the Alices are allowed to 
perform entangled unitaries, this will not enhance the dense coding 
capacity of the state \(\rho^{A_1 A_2 \cdots A_N B}\). 
We will illustrate the case of many Alices in detail for clarity.
However, as long as one considers  unitary encodings, it is clear 
that the Holevo bound is the same for factorised unitaries, and many
Alices are equivalent to a single one with the according dimension.

\section{Holevo-like upper bound on locally accessible information}
\label{sec-RoyalBengal}

The Holevo bound is an upper bound on the accessible information encoded 
in a quantum ensemble that is sent to a \emph{single} receiver.  
This is also an upper bound on the accessible information encoded in a quantum ensemble that is sent to 
\emph{two} receivers, where the receivers are 
allowed to perform only local operations and classical communication (LOCC).
However, in Ref. \cite{Khajuraho}, we have obtained an independent upper bound 
for this situtation. (For a 
lower bound, see Ref. \cite{dum-dum-diga-diga}.) Suppose that a sender encodes 
the classical message \(i\) in the bipartite quantum state \(\rho^{B_1B_2}_i\) with probability \(p_i\), and sends it to two Bobs 
(Bob1 (\(B_1\)) and Bob2 (\(B_2\))).
The tasks of the Bobs is to gather as much information as posssible about \(i\). Let the accessible information in this situation be called ``locally accessible information'', denoted by 
\(I_{acc}^{LOCC}\). It was shown in Ref. \cite{Khajuraho} that 
\beq
I_{acc}^{LOCC} \leq \chi^{LOCC} \equiv S(\overline{\rho}^{B_1}) + S(\overline{\rho}^{B_2}) - \max_{Z=B_1,B_2} p_i S(\rho_i^{Z}),
\eeq
where 
\(\rho_i^{B_1} = \Tr_{B_2} \rho_i^{B_1B_2}\), \(\rho_i^{B_2} = \Tr_{B_1} \rho_i^{B_1B_2}\), \(\overline{\rho}^{Z} = \sum p_i \rho_i^{Z}\), \(Z=B_1, B_2\). 

This bound is not necessarily better than the Holevo bound for all ensembles. For example, for the ensemble formed by the states
\(|00\rangle\), \(|11\rangle\), taken with probability \(\frac{1}{2}\) each, the Holevo bound equals \(1\), while our local 
bound \(\chi^{LOCC}\) is \(2\). 
This, of course, implies 
that the bound \(\chi^{LOCC}\) on \(I_{acc}^{LOCC}\) is asymptotically not attainable in general. 
However, there are important examples for which the local bound (\(\chi^{LOCC}\)) is drastically smaller than 
the global one (\(\chi\)). For example, for the four Bell states \(|\psi^\pm\rangle, |\phi^\pm\rangle \), chosen with 
probabilities \(p_i\) (\(i = 1,2,3,4\)), \(\chi = H(\{p_i\})\), while \(\chi^{LOCC} = 1\). In particular, for equal apriori probabilities, 
the global bound is 2, while the local one is still unity.

\section{Capacity of dense coding with many senders and two receivers}
\label{sec-dui-receivers}

We will now consider the case of dense coding with two receivers. 
Suppose therefore that \(N\) Alices (\(A_1, A_2, \cdots, A_N\)) and 
two Bobs (\(B_1\) and \(B_2\)) share a quantum state \(\rho^{A_1, A_2, \cdots, A_NB_1B_2}\). To send the classical 
information \(i_k\), \(A_k\) performs the unitary operation \(U_{i_k}\), with probability \(p_{i_k}\). Then the Alices send their part of the 
resulting state to the Bobs. For definiteness, let us assume that \(A_1, A_2, \cdots, A_M\) send their parts of the resulting state 
to \(B_1\), while the rest of the Alices send to \(B_2\). Hence the Bobs receive the ensemble 
\(\{p_{\{i\}}, \rho_{\{i\}}\}\), where 
\(p_{\{i\}} = p_{i_1}p_{i_2} \cdots p_{i_N}\), 
\(\rho_{\{i\}} =  U_{\{i\}} \otimes  \Id \otimes \Id \rho^{A_1A_2 \cdots A_N B_1B_2} 
U_{\{i\}}^\dagger \otimes \Id \otimes \Id \), with 
\(U_{\{i\}} = U_{i_1} \otimes U_{i_2} \otimes \cdots \otimes U_{i_N}\).
Let us warn here that the same notation 
\(\rho_{\{i\}}\) was used in the case of a single receiver in Section  \ref{Ujjain}, although the situation there is 
different than this one. 
The aim of  the Bobs is to gather maximal information from the ensemble \(\{p_{\{i\}}, \rho_{\{i\}}\}\) 
about the message string \(\{i\} = \{i_1,i_2, \cdots, i_N\}\), but 
they are restricted to perform only LOCC between themselves. 
The ``one-capacity'' in this case is 
\beq
C^{(1)}_{LOCC}(\rho)= \max_{p_{\{i\}}, U_{\{i\}}} I_{acc}^{LOCC}(\{p_{\{i\}}, \rho_{\{i\}}\}),
\eeq
so that 
\begin{equation}
\label{murlidhar-uncle}
C^{(1)}_{LOCC}(\rho) \leq \max_{p_{\{i\}}, U_{\{i\}}} \chi^{LOCC}(\{p_{\{i\}}, \rho_{\{i\}}\}),
\end{equation}
where the ensemble states \(\rho_{\{i\}}\) in the two above equations is to be considered
in the \(A_1A_2 \cdots A_MB_1 : A_{M+1} A_{M+2} \cdots A_NB_2\) bipartite split, for calculating the 
locally accessible information and its local bound. 
We have 
\begin{equation}
\label{gosto}
\chi^{LOCC}(\{p_{\{i\}}, \rho_{\{i\}}\}) = S(\overline{\rho}^{\textbf{1}}) + S(\overline{\rho}^{\textbf{2}}) 
- \max_{\textbf{Z}=\textbf{1},\textbf{2}} p_{\{i\}} S(\rho_{\{i\}}^{\textbf{Z}}),
\end{equation}
where $\rho_{\{i\}}^{\textbf{1}} = 
\Tr_{A_{M+1}
\cdots A_N B_2} \rho_{\{i\}}^{A_1 \cdots A_N B_1 B_2}$,
$\rho_{\{i\}}^{\textbf{2}} = 
\Tr_{A_1 \cdots A_M B_1} 
\rho_{\{i\}}^{A_1 \cdots A_N B_1 B_2}$, and
$\overline{\rho}^{\textbf{Z}} = \sum p_{\{i\}} \rho_{\{i\}}^{\textbf{Z}}, 
\textbf{Z}=\textbf{1},\textbf{2}$.

The last term on the rhs of Eq. (\ref{gosto}) equals 
\( - \max_{\textbf{Z}= \textbf{1}, \textbf{2}} S(\rho^{\textbf{Z}}) \), for any choice of unitaries and probabilities 
in the maximization of Eq. (\ref{murlidhar-uncle}), where 
\begin{eqnarray}
\label{kalipur}
\rho^{\textbf{1}} = \Tr_{A_{M+1}A_{M+2} \cdots A_NB_2} \rho, 
\qquad\rho^{\textbf{2}} = \Tr_{A_{1}A_{2} \cdots A_MB_1} \rho.
\end{eqnarray}

Next, note that the maximization in Eq. (\ref{murlidhar-uncle}) of the first two terms on the rhs of Eq. (\ref{gosto})
can be independently performed. For example, the maximization of 
\(S(\overline{\rho}^{\textbf{1}})\) can be performed solely on the probabilities
\(p_1, p_2, \cdots, p_M\), and the unitaries \(U_1, U_2, \cdots, U_M\)
and can be done as 
in Section \ref{Ujjain}.
Similar considerations hold for the maximization of 
\(S(\overline{\rho}^{\textbf{2}})\)  over the probabilities 
\(p_{M+1}, p_{M+2}, \cdots, p_N\), and the unitaries \(U_{M+1}, U_{M+2}, \cdots, U_N\).
So finally, we have 
\begin{eqnarray}
\label{radha-kakima}
C^{(1)}_{LOCC}(\rho) \leq \log_2d_{A_1} + \cdots+ \log_2d_{A_N} + S(\rho^{B_1}) + S(\rho^{B_2}) -  
\max_{\textbf{Z}= \textbf{1}, \textbf{2}} S(\rho^{\textbf{Z}}).
\end{eqnarray}

For unitary encoding, the rhs of 
Eq. (\ref{gosto}) is additive, and
so the asymptotic capacity of distributed dense coding is also bounded by the same quantity:
\begin{eqnarray}
\label{la-rambla}
C_{LOCC}(\rho) \leq \log_2d_{A_1} + \cdots + \log_2d_{A_N}  
+ S(\rho^{B_1}) + S(\rho^{B_2}) -  \max_{\textbf{Z}= 
\textbf{1}, \textbf{2}} S(\rho^{\textbf{Z}}).
\end{eqnarray}
The partition in Eq. (\ref{kalipur}) corresponds to the partition 
in two Bobs' states after they received the states \(\rho_{\{i\}}\). 
In general, the local capacities of the state depend 
on this partition.



\section{A classification of multiparty states by their dense-codeability}
\label{sec-nyapla}

A simple lower bound on \(C_{LOCC}\) can be obtained by considering the case when the two Bobs do not use communication,
whereby the two channels (one from the first \(M\) Alices to \(B_1\) and 
the other from the next \(N-M\) Alices to \(B_2\)) are
independent, and so the capacities add. Let us denote the capacity
without communication as $C_{LO}$, and thus have
\beq
C_{LOCC}(\rho) \geq  C_{LO}(\rho) = 
C(\rho^{\textbf{1}}) + C(\rho^{\textbf{2}}),
\eeq
where 
\(C(\rho)\) is given by Eq. (\ref{kanishka}), and \(\rho^{\textbf{1}}\) and \(\rho^{\textbf{2}}\) are defined in Eq. (\ref{kalipur}).
If the Bobs are together, and are allowed to perform global measurements, 
then the capacity is given by using Eq. (\ref{kanishka}). This capacity is also an 
upper bound of \(C_{LOCC}\). Therefore, 
\begin{eqnarray}
\label{bashirhat}
C_{LOCC}(\rho ) \leq \log_2 d_{A_1}   + \log_2 d_{A_2} + \cdots + 
\log_2 d_{A_{N}} + S(\rho^{B_1B_2}) -  S(\rho)  = C_G( \rho). 
\end{eqnarray}

The rhs of the above inequality (\ref{bashirhat}) is precisely the dense coding capacity of the state \(\rho\), when the two receivers 
are together, and hence
are allowed to perform global measurements. We have denoted this quantity by \(C_G (\rho)\).
With the help of the quantities \(C_G\), \(C_{LOCC}\), \(C_{LO}\), and the relations between them, 
multipartite states can be classified according to their usefulness for dense-coding. 
Consider therefore the \(N+2\)-partite state \(\rho^{A_1A_2 \cdots A_N B_1 B_2}\), and 
consider first the bipartite split \(A_1A_2 \cdots A_N :  B_1 B_2\). This is the senders to receivers bipartite split
in the distributed dense coding scenario. In this bipartite split, the 
usual classification is into four classes: Separable states (S), 
bound entangled states with positive 
partial transpose (PBE) \cite{ref-Gdansk}, bound entangled states with nonpositive partial transpose (NBE)  (if 
existing) \cite{nbe}, and distillable states. 
As shown in Sec. 
\ref{sec-Leibnizhaus}, bound entangled states (both PBE and NBE), as well as separable states are not useful 
for dense coding. Thus only distillable states can be useful. However, not 
all distillable states can be used. For example,
even for \(2 \otimes 2\) states,  the Werner state \cite{Werner}
\beq
\rho_p = p |\psi^- \rangle \langle \psi^- | + (1 -p ) \frac{I \otimes I}{4}
\eeq
is distillable  when  \(p \geq \frac{1}{3}\). 
But using Eq. (\ref{ek-ekcapacity}), one can see that
the state \(\rho_p\) is good for dense coding only for 
$p \geq 0.7476$. 
Going back to our multiparty state \(\rho^{A_1A_2 \cdots A_N B_1 B_2}\)
in the bipartite split \(A_1A_2 \cdots A_N :  B_1 B_2\), the distillable states are divided into two categories: 
Ones which are globally dense-codeable, and 
ones which are not. The globally dense-codeable (G-DC) states are those which can be useful for 
dense coding when the two Bobs are at the same location. Therefore they are precisely those 
for which \(C_G > \log_2 d_{A_1} + \log_2 d_{A_2} + \cdots + \log_2 d_{A_N} \), i.e. for which
\(S(\rho^{B_1B_2}) > S(\rho)\). 
The states which are distillable in the \(A_1A_2 \cdots A_N :  B_1 B_2\) split, and yet are not useful for 
dense coding are denoted by D. 

\begin{figure}[h]
\begin{center}
  \includegraphics[width=10cm]{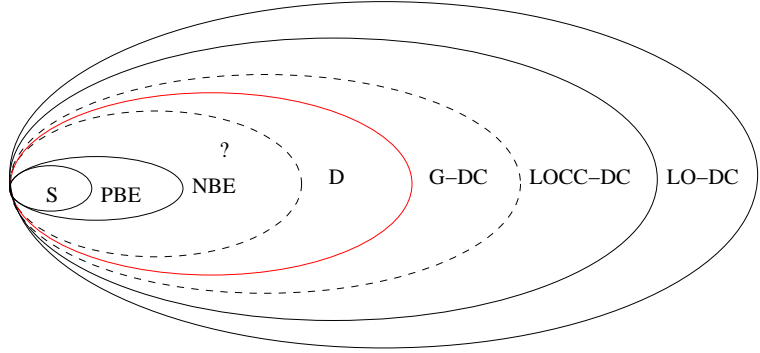}
\end{center}    
\caption{Classification of {\em multipartite} quantum states,
according to their usefulness for dense coding with more than one receiver. 
Notice that the labels classify only the states in the shell and not in the 
whole set (ellipse). Separable, bound entangled states with 
positive partial transpose, bound entangled states with nonpositive 
partial transpose (if existing), distillable but not useful for dense coding 
respectively are denoted as
S, PBE, NBE, D. 
In the bipartite case, there is just one more shell, consisting of 
 states which are distillable and can be used for dense coding. These
states are in the shell G-DC. 
In the multiparty case, there also exist shells which contain states that are good for G-DC but not good for LOCC-DC.
Similarly, the
 shell denoted as LOCC-DC contain states who are 
useful for LOCC-DC but not for LO-DC, as explained in the text. Also there
are states which are good for dense coding even without communication (LO-DC). 
As discussed in the text, all shells are non-empty and of nonzero measure.
Borders between sets that are not known to be convex are drawn as dashed lines.
}
 \label{figure1}
\end{figure}
Although the classification above into S, PBE, NBE, D, and G-DC was considered for a 
multiparty state, this is essentially the classification for bipartite states. 
This classification is summarized in
Fig. \ref{figure1}, where for 
the bipartite case, only the classes S, PBE, NBE, D, and G-DC are meaningful. 
The multiparty case offers a much richer classification: the states 
\(\rho^{A_1 \cdots A_N B_1 B_2}\) that are 
distillable in the \(A_1 \cdots A_N :  B_1 B_2\) split,
can in this case be divided into the following four classes:
\begin{enumerate}
\item LO-DC class: This class contains states that can be used for dense coding even when the Bobs are separated and 
they do not even communicate classically. 
Precisely, they are those for which 
\begin{equation}
C_{LO} > \log_2 d_{A_1}   + \log_2 d_{A_2} + \cdots + \log_2 d_{A_{N}},
\end{equation} 
i.e. 
for which \(S(\rho^{B_1}) + S(\rho^{B_2}) > S(\rho^{A_1A_2 \cdots A_M B_1}) + S(\rho^{A_{M+1}A_{M+2} \cdots A_N B_2})\). 

\item LOCC-DC class: This class contains states that are useful for dense coding when the two Bobs are separated, 
but they 
are allowed to communicate classically. So, these are states for which 
\begin{equation}
C_{LOCC} >   \log_2 d_{A_1}   + \log_2 d_{A_2} + \cdots + \log_2 d_{A_{N}}.
\end{equation}
Moreover, we require 
that the states in the LOCC-DC class to be not LO-DC. 

\item G-DC class: This class contains states that are useful for dense coding when 
the two Bobs are at the same location. Therefore, for these states
\begin{equation}
C_{G} >\log_2 d_{A_1}  
 + \log_2 d_{A_2} + \cdots + \log_2 d_{A_{N}}.
\end{equation}
 Again we also require that the states in the 
G-DC class are not LOCC-DC.

\item D class: The final class contains the states that are distillable in the 
\(A_1A_2 \cdots A_N : B_1B_2\) split, but not G-DC:
\begin{equation}
C_{G} \leq \log_2 d_{A_1}  
 + \log_2 d_{A_2} + \cdots + \log_2 d_{A_{N}}.
\end{equation}

\end{enumerate}

\subsection{Examples}

We will now give examples for all the above classes. 
We have already shown that the Werner states provide examples of states which are distillable, and yet are 
not useful for dense coding. Similar examples exist for GHZ states \cite{GHZ} admixed with white noise:
\(p|\mbox{GHZ}\rangle \langle \mbox{GHZ}| + (1-p) I^{\otimes n}/2^n\) where 
\(|\mbox{GHZ}\rangle = (\left|0\right\rangle^{\otimes n} + \left|1\right\rangle^{\otimes n})/\sqrt{2}\). 

There also exist states by which dense coding is possible only when the receivers (\(B_1\) and \(B_2\)) are together. 
An example of such a state is 
\beq
\frac{1}{2} \left( \left|0000\right\rangle + \left|0101\right\rangle +
\left|1000\right\rangle + \left|1110\right\rangle \right)
\eeq
from Ref. \cite{Frank123}. 
Here the first two parties are senders and they perform the unitary operations. Then the first party sends her part of the 
multiparty state to the third party, 
while the second one sends her part to the fourth party. 
For this state, \(C_G > \log_2 d_{A_1}   + \log_2 d_{A_2} + \cdots + \log_2 d_{A_{N}}\) 
but the upper bound of \(C_{LOCC}\) in Eq. (\ref{la-rambla}) is less than 
\(\log_2 d_{A_1} + \log_2 d_{A_2} + \cdots + \log_2 d_{A_N} \) (with 
\(N = 2\) and \(d_{A_1} = d_{A_2} = 2\)). 

Let us now consider the four-qubit GHZ state, namely 
$(\left|0000\right\rangle + \left|1111\right\rangle)/\sqrt{2}$.
We will now show that this state is useful 
for dense coding,
even when the receivers are restricted only to LOCC operations. 
However the capacity \(C_{LO}\) of the GHZ state
 is vanishing, since its two-particle local 
density matrices are separable. 
Suppose therefore that the four-qubit GHZ state (ignoring normalization)
 \(\left|\mbox{GHZ}_4 \right\rangle^{A_1 A_2 B_1 B_2} = \left|0000\right\rangle + \left|1111\right\rangle\)  
is shared by four far-apart partners \(A_1\), \(A_2\), \(B_1\), and 
\(B_2\).  \(A_1\), \(A_2\)  perform the unitary operations 
\(I\), \(\sigma_x\), \(\sigma_y\),  \(\sigma_z\)  (\(\sigma_x\), \(\sigma_y\), \(\sigma_z\) 
are the Pauli matrices), with equal probabilities. Then
 \(A_1\)  sends her qubit to \(B_1\) and \(A_2\) to \(B_2\).  
\(B_1\)
and \(B_2\) then  share the states \(\{\left|\psi_i\right\rangle\}_{i=1}^{8}\), of the 
eight orthogonal states with equal probabilities, given by
\begin{eqnarray}
\begin{array}{rcl}
\left|\psi_{1,2}\right\rangle & = & \left|00\right\rangle^{B_1}\left|00\right\rangle^{B_2} \pm \left|11\right\rangle^{B_1} \left|11\right\rangle^{B_2}, \\
\left|\psi_{3,4}\right\rangle & = & \left|00\right\rangle^{B_1}\left|10\right\rangle^{B_2} \pm \left|11\right\rangle^{B_1} \left|01\right\rangle^{B_2},\\
\left|\psi_{5,6}\right\rangle & = & \left|10\right\rangle^{B_1}\left|00\right\rangle^{B_2} \pm \left|01\right\rangle^{B_1} \left|11\right\rangle^{B_2},\\
\left|\psi_{7,8}\right\rangle & = & \left|10\right\rangle^{B_1}\left|10\right\rangle^{B_2} \pm \left|01\right\rangle^{B_1} \left|01\right\rangle^{B_2},
\end{array}
\end{eqnarray}
where the smaller index on the lhs corresponds to
the upper sign on the rhs. 
For the decoding (by LOCC between \(B_1\) and \(B_2\)), 
\(B_1\) begins by making a measurement with the projectors
\( P_0 =\left|00\right\rangle \left\langle 00 \right| + \left|11\right\rangle \left\langle 11 \right|\), 
\(P_1= \left|01\right\rangle \left\langle 01 \right| + \left|10\right\rangle \left\langle 10 \right|\) 
and communicates the result to \(B_2\).  If \(P_0\) (\(P_1\)) clicks, then they know that the  
state  is among \(\left|\psi_i\right\rangle, i \in \{1,2,3,4\}\) (\(\left|\psi_i\right\rangle, i \in \{5,6,7,8\}\)).
Now \(B_2\) performs a measurement with the same projectors \(P_0\), \(P_1\). 
Depending on the outcome, they know that the state they share is either 
\(\left|\psi_{1,2}\right\rangle\), or 
\(\left|\psi_{3,4}\right\rangle\), or 
\(\left|\psi_{5,6}\right\rangle\), or
\(\left|\psi_{7,8}\right\rangle\).
Note that none of the above measurements 
disturbs the shared state. 
Lastly, performing a measurement
in \(\{\left|00\right\rangle \pm \left|11\right\rangle \}\) or \(\{\left|01\right\rangle \pm \left|10\right\rangle \}\) basis (depending on the outcomes in 
the previous measurements) by both the Bobs on their respective sides, 
will help them to locally distinguish the state perfectly. The above protocol for dense coding
and the upper bound in Eq. (\ref{la-rambla}),
imply that \(C_{LOCC} = 3\), for the four-qubit GHZ, which is therefore LOCC-DC. 

An example for which the capacity \(C_{LO}\) is non-zero
is \(|\psi^-\rangle^{A_1B_1} \otimes |\psi^-\rangle^{A_2B_2} \).
It is actually non-zero 
for tensor product of any two bipartite states \(\rho^{A_1B_1}\) and \(\rho^{A_2B_2}\), 
which are independently 
useful in dense coding with a single sender and a single receiver, i.e. 
for which \(C(\rho^{A_1B_1}) + C(\rho^{A_2B_2}) > \log_2 d_{A_1} + \log_2 d_{A_2}\). 

The boundary between LO-DC and LOCC-DC states is given by 
\(C_{LO} = \log_2 d_{A_1} + \log_2 d_{A_2} + \cdots + \log_2 d_{A_N}\). For four qubit states, with two 
senders and two receivers, the boundary is given by \(C_{LO} = 2\). Now for the state 
\(|\psi^-\rangle \otimes |\psi^-\rangle \), we have \(C_{LO} = 4\), so that it is far from the boundary. 
(It actually possesses the maximal dense coding capacity reachable by any four qubit state with two senders and 
two receivers.) Consequently, by continuity, one can argue that this state will remain away from the boundary 
even after admixture of sufficiently small amount of noise.  This implies that the LO-DC class has a 
nonzero measure. A similar way of arguing is possible for all other examples corresponding to the 
different classes considered above. In particular, the 
LOCC-DC class can be proven to be of nonzero measure by considering 
noise admixture to the four qubit GHZ state.

\subsection{Convexity of the classes}

Now we consider the question of convexity of the boundaries between the shells
considered in Fig. \ref{figure1}. 
Separable states form a convex set. 
So do the states with positive partial transpose (PPT), i.e. separable and PPT bound entangled states, 
since adding two PPT states
never gives a state whose partial transpose is non-positive. 
It was shown in Ref. \cite{SST} 
that the boundary between the NBE and D 
shells is not convex, if a certain NBE state exists (see also \cite{graph}). 
The D to G-DC boundary 
is convex since the conditional entropy \(S(\rho^{AB}) - S(\rho^B) \) 
is a concave function \cite{Wehrl}. 
The LOCC-DC
to LO-DC boundary is convex due to the same reason, as 
it is the sum of two convex quantities, viz. the two single receiver capacities.
However the  convexity of the G-DC  to LOCC-DC boundary is not known.

\section{Discussion}
\label{disco}

In this paper, we have introduced dense coding protocols for multipartite 
states where all the parties are far apart. 
We have considered two types of schemes: 
one with several senders and a single receiver, and another with several senders 
and two receivers who are allowed to perform only local operations. 
In the first case, we found the exact capacity of the channel while 
in the latter case, we provide a useful upper bound. 
In the latter case, we have also shown that the GHZ state 
achieves the upper bound. These two protocols help us to classify
multipartite states from the point of view of usefulness for dense coding. 
In the bipartite case, this classification is complete. We know that 
separable states as well as bound entangled states  are not useful for dense 
coding,
while highly distillable states are good for it. There exist some distillable states which are 
not useful for dense coding. 
However in the multipartite situation, several  questions remain open, 
both for 
one and two receiver(s). 
For example, we do not know whether multipartite bound entangled states are useful in such schemes. 
Let us consider the ``unlockable'' bound entangled state 
\beq
\rho_{S} = \frac{1}{4} \sum_i |\psi_i\rangle \langle \psi_i| \otimes |\psi_i\rangle \langle \psi_i| 
\eeq
of Ref. \cite{Smolin},
where the \(|\psi_i\rangle\)s are the Bell states.
Let \(\rho_S\) be
shared between \(A_1\), \(A_2\), \(B_1\), \(B_2\). 
\(\rho_S\) is separable in all two party by two party splittings, although
it has one bit of entanglement in all 
one party by three party splittings.
One can check by using Eq. (\ref{bashirhat}) that the \(C_G\) of \(\rho_S\)
is not greater than 3 bits, but exactly
equal to 3 bits, when \(A_1\), \(A_2\), \(B_1\) are senders and \(B_2\) is the receiver.  
Since all its  two party by two party splittings are separable, 
it is clear that it will never be 
useful for dense coding with two receivers. We have also checked our formulas for other bound entangled states, 
e.g. the bound entangled states 
formed from the unextendible product bases \cite{UPB}, and they are not
useful for dense coding either. 

In this paper, we have considered distributed communication protocols, where 
the senders are only 
allowed to perform unitary operations. This case is more interesting from the perspective of 
a real implementation. However the 
Holevo-like upper bound \cite{Khajuraho}  on accessible information holds for any
encoding (as well as decoding) operation. 
So, it is also interesting to consider general encoding protocols,
and obtain  upper bounds on distributed communication rates in this case.  
For the latest development of this general case in a situation, where there is only 
a single sender and a single receiver, see e.g. \cite{MarieCurie,Debu}. 
In this paper, it is always assumed that
the transmission channel is noiseless, even if the shared states 
that we use as our resource may be noisy. 
Even in the case of such noiseless channels, we show that the
states that we require in such communication 
are highly entangled. 
It would be interesting to study the dense coding capacity of noisy states 
in the realistic case of noisy channels.

\section*{Acknowledgments}
We thank Alexander Holevo and Mario Ziman  for valuable comments. 
We acknowledge support from the Deutsche Forschungsgemeinschaft (SFB 407, SPP 1078, SPP 1116, 436POL), 
the Alexander von Humboldt Foundation, the EC Program 
QUPRODIS, the ESF Program QUDEDIS, and EU IP SCALA.


\begin{thebibliography}{99}

\bibitem{BBCJPW} C.H. Bennett {\it et al.},
Phys. Rev. Lett. \textbf{70}, 1895 (1993).

\bibitem{nischintapur} C.H. Bennett and S.J. Wiesner, 
Phys. Rev. Lett. \textbf{69}, 2881 (1992).

\bibitem{Bennett-ek} C.H. Bennett {\it et al.},
Phys. Rev. Lett. \textbf{83}, 3081 (1999).

\bibitem{Bennett-dui} C.H. Bennett {\it et al.}, Entanglement-assisted capacity of a quantum channel and the reverse Shannon theorem, 
quant-ph/0106052.


\bibitem{MarieCurie} M. Horodecki {\it et al.}, 
Q. Inf. and Comput. \textbf{1}, 70 (2001).

\bibitem{Debu} A. Winter, J. Math. Phys. \textbf{43}, 4341 (2002). 


\bibitem{ek} S. Bose, M.B. Plenio, and V. Vedral, Mixed state dense coding and its relation to entanglement measures, quant-ph/9810025.

\bibitem{dui} T. Hiroshima,  J. Phys. A: Math. Gen. \textbf{34},  6907 (2001).


\bibitem{tin} M. Ziman and V. Bu{\v z}ek, Phys. Rev. A \textbf{67}, 042321 (2003).

\bibitem{char} X.S. Liu {\it et al.}, Phys. Rev. A \textbf{65}, 022304 (2002).

\bibitem{TajMahal}  D. Bru{\ss} {\it et al.}, 
Phys. Rev. Lett. \textbf{93}, 210501 (2004).

\bibitem {ref-halum}
J.P. Gordon, in \emph{Proc. Int. School Phys. ``Enrico Fermi,
Course XXXI''}, ed. P.A. Miles, pp. 156 (Academic Press, NY 1964); 
L.B. Levitin, in \emph{Proc. VI National Conf. Inf. Theory, Tashkent}, pp. 111
(1969); A.S. Holevo, Probl. Pereda. Inf. \textbf{9}, 3 1973 [Probl. Inf.
Transm. \textbf{9}, 110 (1973)].


\bibitem{Khajuraho} P. Badzi{\c a}g {\it et al.}, Phys. Rev. Lett. 
\textbf{91}, 117901 (2003).

\bibitem{Sundarban} M. Horodecki, A. Sen(De), and U. Sen, Quantification of quantum correlation of ensemble of states, quant-ph/0310100. 

\bibitem{Chennai}  T. M. Cover and J. A. Thomas, \emph{Elements of Information Theory} (Wiley, 1991).

\bibitem{Utpakhi} R. Josza, D. Robb, and W.K. Wotters, Phys. Rev. A, \textbf{49}, 668 (1994).

\bibitem{Rajabazar1} B. Schumacher {\it et al.},
Phys. Rev. Lett. \textbf{76}, 3452 (1996).

\bibitem{Rajabazar2} M Horodecki {\it et al.},  
Phys. Rev. Lett. \textbf{93}, 170503 (2004) .

\bibitem{babarey} B. Schumacher and M.D. Westmoreland, Phys. Rev. A \textbf{56}, 131 (1997); A.S. Holevo, IEEE Trans. Inf. Theory \textbf{44}, 269 (1998).

\bibitem{ref-Gdansk} M. Horodecki, P. Horodecki, and  R. Horodecki, Phys. Rev. Lett. \textbf{80}, 5239 (1998).

\bibitem{reduction} M. Horodecki and P. Horodecki, Phys. Rev. A \textbf{59}, 4206 (1999); N.J. Cerf, C. Adami, and R.M. Gingrich, \emph{ibid}.
\textbf{60}, 898 (1999).

\bibitem{Wolf} K.G.H. Vollbrecht and M.M. Wolf, Conditional entropies and their relation to entanglement criteria, 
quant-ph/0202058.

\bibitem{etao-dewa-uchit} T. Hiroshima, Phys. Rev. Lett. \textbf{91}, 057902 (2003).


\bibitem{dum-dum-diga-diga}  A. Sen(De), U. Sen, and M. Lewenstein, Lower Bound on Locally Accessible Information: Local Subentropy, 
quant-ph/0505137.

\bibitem{nbe}D. P. DiVincenzo {\it et al.}, Phys. Rev. A  \textbf{61}, 
062312 (2000); W. D\" ur {\it et al.}, Phys. Rev. A \textbf{61}, 062313 (2000).

\bibitem{Werner} R.F. Werner, Phys. Rev. A \textbf{40}, 4277 (1989).

\bibitem{GHZ} D.M. Greenberger, M.A. Horne, and A. Zeilinger, in \emph{Bell's
 Theorem, Quantum Theory, and Conceptions of the Universe}, ed. M. Kafatos,
 (Kluwer, Dordrecht, 1989).

\bibitem{Frank123}F. Verstraete {\it et al.},   
Phys. Rev. A \textbf{65}, 052112 (2002).

\bibitem{SST} P.W.  Shor, J.A. Smolin, and B.M. Terhal,
Phys. Rev. Lett. \textbf{86}, 2681 (2001).

\bibitem{graph}  T. Eggeling {\it et al.}, 
Phys. Rev. Lett. \textbf{87}, 257902 (2001).

\bibitem{Wehrl} A. Wehrl, Rev. Mod. Phys. \textbf{50}, 221 (1978).

\bibitem{Smolin} J.A. Smolin, Phys. Rev. A \textbf{63}, 032306 (2001).

\bibitem{UPB} C.H. Bennett {\it et al.}, Phys. Rev. Lett. \textbf{82}, 5385  
(1999);  D.P. DiVincenzo {\it et al.}, 
Comm. Math. Phys. \textbf{238},  379 (2003).

 

\end{thebibliography}
\end{document}